# Homomorphic Encryption: Theory & Application

Jaydip Sen

*Department of Computer Science, National Institute of Science & Technology*
*Odisha, INDIA*

## 1. Introduction

The demand for privacy of digital data and of algorithms for handling more complex structures have increased exponentially over the last decade. This goes in parallel with the growth in communication networks and their devices and their increasing capabilities (Sen, 2013a; Sen, 2012d; Sen, 2012c; Bandyopadhyay & Sen, 2011; Sen, 2011b; Sen, 2011c; Sen, 2010c). At the same time, these devices and networks are subject to a great variety of attacks involving manipulation and destruction of data and theft of sensitive information (Sen, 2013b; Sen, 2012e; Sen, 2011d; Sen, 2010d; Sen et al., 2007a; Sen et al., 2007b; Sen et al., 2007c; Sen et al., 2006). For storing and accessing data securely, current technology provides several methods of guaranteeing privacy such as data encryption and usage of tamper-resistant hardwares. However, the critical problem arises when there is a requirement for computing (publicly) with private data or to modify functions or algorithms in such a way that they are still executable while their privacy is ensured. This is where homomorphic cryptosystems can be used since these systems enable computations with encrypted data.

In 1978 Rivest et al. (Rivest et al, 1978a) first investigated the design of a homomorphic encryption scheme. Unfortunately, their *privacy homomorphism* was broken a couple of years later by Brickell and Yacobi (Brickell & Yacobi, 1987). The question rose again in 1991 when Feigenbaum and Merritt (Feigenbaum & Merritt, 1991) raised an important question: *is there an encryption function (E) such that both E(x + y) and E(x.y) are easy to compute from E(x) and E(y)?* Essentially, the question is intended to investigate whether there is any algebraically homomorphic encryption scheme that can be designed. Unfortunately, there has been a very little progress in determining whether such encryption schemes exist that are efficient and secure until 2009 when Craig Gentry, in his seminal paper, theoretically demonstrated the possibility of construction such an encryption system (Gentry, 2009). In this chapter, we will discuss various aspects of homomorphic encryption schemes – their definitions, requirements, applications, formal constructions, and the limitations of the current homomorphic encryption schemes. We will also briefly discuss some of the emerging trends in research in this field of computer science.

The chapter is organized as follows. In Section 2, we provide some basic and fundamental information on cryptography and various types of encryption schemes. Section 3 presents a formal discussion on homomorphic encryption schemes and discusses their various features. In Section 4, we discuss some of the most well-known and classical homomorphic encryption schemes in the literature. Section 5 provides a brief presentation on various properties and applications of homomorphic cryptosystems. Section 6 presents a discussion on fully homomorphic encryption schemes which are the most powerful encryption schemes for providing a framework for computing over encrypted data. Finally, Section 7 concludes the chapter while outlining a number of research directions and emerging trends in this exciting filed of computation which has a tremendous potential of finding applications in the real-world deployments.

## 2. Fundamentals of Cryptography

In this Section, we will recall some important concepts on encryption schemes. For more detailed information, the reader may refer to (Menezes et al., 1997; Van Tilborg, 2011). Encryption schemes are designed to preserve confidentiality. The security of encryption schemes must not rely on the obfuscation of their codes, but it should only be based on the secrecy of the key used in the encryption process. Encryption schemes are broadly of two types: *symmetric* and *asymmetric* encryption schemes. In the following, we present a very brief discussion on each of these schemes.

**Symmetric encryption schemes**: In these schemes, the sender and the receiver agree on the key they will use before establishing any secure communication session. Therefore, it is not possible for two persons who never met before to use such schemes directly. This also implies that in order to communicate with different persons, we must have a different key for each people. Requirement of large number of keys in these schemes make their key generation and management relatively more complex operations. However, symmetric schemes present the advantage of being very fast and they are used in applications where speed of execution is a paramount requirement. Among the existing symmetric encryption systems, AES (Daemen & Rijmen, 2000; Daemen & Rijmen, 2002), One-Time Pad (Vernam, 1926) and Snow (Ekdahl & Johansson, 2002) are very popular.

**Asymmetric encryption schemes:** In these schemes, every participant has a pair of keys-private and public. While the private key of a person is known to only her, the public key of each participant is known to everyone in the group. Such schemes are more secure than their symmetric counterparts and they don't need any prior agreement between the communicating parties on a common key before establishing a session of communication. RSA (Rivest et al., 1978b) and ElGamal (ElGamal, 1985) are two most popular asymmetric encryption systems.

**Security of encryption schemes:** Security of encryption schemes was first formalized by Shannon (Shannon, 1949). In his seminal paper, Shannon first introduced the notion of perfect secrecy/unconditional secrecy, which characterizes encryption schemes for which the knowledge of a ciphertext does not give any information about the corresponding plaintext and the encryption key. Shannon also proved that One-Time Pad (Vernam, 1926) encryption scheme is perfectly secure under certain conditions. However, no other encryption scheme has been proved to be unconditionally secure. For asymmetric schemes, we can rely on their mathematical structures to estimate their security strength in a formal way. These schemes are based on some well-identified mathematical problems which are hard to solve in general, but easy to solve for the one who knows the trapdoor – i.e., the owner of the keys. However, the estimation of the security level of these schemes may not be always correct due to several reasons. First, there may be other ways to break the system than solving the mathematical problems on which these schemes are based (Ajtai & Dwork, 1997; Nguyen & Stern, 1999). Second, most of the security proofs are performed in an idealized model called *random oracle model*, in which involved primitives, for example, hash functions, are considered truly random. This model has allowed the study of the security

Homomorphic Encryption: Theory and Application

level of numerous asymmetric ciphers. However, we are now able to perform proofs in a more realistic model called *standard model* (Canetti et al., 1998; Paillier, 2007). This model eliminates some of the unrealistic assumptions in the random oracle model and makes the security analysis of cryptographic schemes more practical.

Usually, to evaluate the attack capacity of an adversary, we distinguish among several contexts (Diffie & Hellman, 1976): *cipher-text only attacks* (where the adversary has access only to some ciphertexts), *known-plaintext attacks* (where the adversary has access to some pairs of plaintext messages and their corresponding ciphertexts), *chosen-plaintext attacks* (the adversary has access to a decryption oracle that behaves like a black-box and takes a ciphertext as its input and outputs the corresponding plaintexts). The first context is the most frequent in real-world since it can happen when some adversary eavesdrops on a communication channel. The other cases may seem difficult to achieve, and may arise when the adversary is in a more powerful position; he may, for example, have stolen some plaintexts or an encryption engine. The *chosen* one exists in adaptive versions, where the opponents can wait for a computation result before choosing the next input (Fontaine & Galand, 2007).

**Probabilistic encryption:** Almost all the well-known cryptosystems are *deterministic*. This means that for a fixed encryption key, a given plaintext will always be encrypted into the same ciphertext under these systems. However, this may lead to some security problems. RSA scheme is a good example for explaining this point. Let us consider the following points with reference to the RSA cryptosystem:

- A particular plaintext may be encrypted in a too much structured way. With RSA, messages 0 and 1 are always encrypted as 0 and 1, respectively.
- It may be easy to compute some partial information about the plaintext: with RSA, the ciphertext $c$ leaks one bit of information about the plaintext $m$, namely, the so called Jacobi symbol (Fontaine & Galand, 2007).
- When using a deterministic encryption scheme, it is easy to detect when the same message is sent twice while processed with the same key.

In view of the problems stated above, we prefer encryption schemes to be probabilistic. In case of symmetric schemes, we introduce a random vector in the encryption process (e.g., in the pseudo-random generator for stream ciphers, or in the operating mode for block ciphers) – generally called *initial vector* (IV). This vector may be public and it may be transmitted in a clear-text form. However, the IV must be changed every time we encrypt a message. In case of asymmetric ciphers, the security analysis is more mathematical and formal, and we want the randomized schemes to remain analyzable in the same way as the deterministic schemes. Researchers have proposed some models to randomize the existing deterministic schemes, as the *optimal asymmetric encryption padding* (OAEP) for RSA (or any scheme that is based on a trapdoor one-way permutation) (Bellare & Rogaway, 1995). In the literature, researchers have also proposed some other randomized schemes (ElGamal, 1985; Goldwasser & Micali, 1982; Blum & Goldwasser, 1985).

A simple consequence of this requirement of the encryption schemes to be preferably probabilistic appears in the phenomenon called *expansion*. Since for a plaintext we require the existence of several possible ciphertexts, the number of ciphertexts is greater than the number of possible plaintexts. This means the ciphertexts cannot be as short as the plaintexts; they have to be strictly longer. The ratio of the length of the ciphertext and the corresponding plaintext (in bits) is called expansion. The value of this parameter is of paramount importance in determining *security and efficiency tradeoff* of a probabilistic encryption scheme. In Paillier's scheme, an efficient probabilistic encryption mechanism has been proposed with the value of expansion less than 2 (Paillier, 1997). We will see the significance of expansion in other homomorphic encryption systems in the subsequent sections of this chapter.

## 3. Homomorphic Encryption Schemes

During the last few years, homomorphic encryption schemes have been studied extensively since they have become more and more important in many different cryptographic protocols such as, e.g., voting protocols. In this Section, we introduce homomorphic cryptosystems in three steps: *what*, *how* and *why* that reflects the main aspects of this interesting encryption technique. We start by defining homomorphic cryptosystems and algebraically homomorphic cryptosystems. Then we develop a method to construct algebraically homomorphic schemes given special homomorphic schemes. Finally, we describe applications of homomorphic schemes.

**Definition:** Let the message space $(M, o)$ be a finite (semi-)group, and let σ be the security parameter. A *homomorphic public-key encryption scheme* (or *homomorphic cryptosystem*) on $M$ is a quadruple $(K, E, D, A)$ of probabilistic, expected polynomial time algorithms, satisfying the following functionalities:

- **Key Generation:** On input $1^o$ the algorithm $K$ outputs an encryption/decryption key pair $(k_e, k_d) = k \in K$, where $K$ denotes the key space.
- **Encryption:** On inputs $1^o$, $k_e$, and an element $m \in M$ the encryption algorithm E outputs a ciphertext $c \in C$, where $C$ denotes the ciphertext space.
- **Decryption:** The decryption algorithm $D$ is deterministic. On inputs $1^o$, $k$, and an element $c \in C$ it outputs an element in the message space $M$ so that for all $m \in M$ it holds: if $c = E(11^\sigma, k_e, m)$ then $Prob[D(1^\sigma, k, c) \neq m]$ is negligible, i.e., it holds that $Prob[D(1^\sigma, k, c) \neq m] \leq 2^{-\sigma}$.
- **Homomorphic Property:** $A$ is an algorithm that on inputs $1^\sigma, k_e$, and elements $c_1, c_2 \in C$ outputs an element $c_3 \in C$ so that for all $m_1, m_2 \in M$ it holds: if $m_3 = m_1 \, o \, m_2$ and $c_1 = E(1^\sigma, k_e, m_1)$, and $c_2 = E(1^\sigma, k_e, m_2)$, then $Prob[D(A(1^\sigma, k_e, c_1, c_2))] \neq m_3]$ is negligible.

Informally speaking, a homomorphic cryptosystem is a cryptosystem with the additional property that there exists an efficient algorithm to compute an encryption of the sum or the product, of two messages given the public key and the encryptions of the messages but not the messages themselves.

If $M$ is an additive (semi-)group, then the scheme is called *additively homomorphic* and the algorithms $A$ is called *Add* Otherwise, the scheme is called *multiplicatively homomorphic* and the algorithm $A$ is called *Mult*.

With respect to the aforementioned definitions, the following points are worth noticing:

Homomorphic Encryption: Theory and Application

- For a homomorphic encryption scheme to be efficient, it is crucial to make sure that the size of the ciphertexts remains polynomially bounded in the security parameter σ during repeated computations.
- The security aspects, definitions, and models of homomorphic cryptosystems are the same as those for other cryptosystems.

If the encryption algorithm $E$ gets as additional input a uniform random number $r$ of a set $\mathbb{Z}$, the encryption scheme is called *probabilistic*, otherwise, it is called *deterministic*. Hence, if a cryptosystem is probabilistic, there belong several different ciphertexts to one message depending on the random number $r \in \mathbb{Z}$. But note that as before the decryption algorithm remains deterministic, i.e., there is just one message belonging to a given ciphertext. Furthermore, in a probabilistic, homomorphic cryptosystem the algorithm $A$ should be probabilistic too to hide the input ciphertext. For instance, this can be realized by applying a *blinding algorithm* on a (deterministic) computation of the encryption of the product and of the sum respectively.

**Notations:** In the following, we will omit the security parameter σ and the public key in the description of the algorithms. We will write $E_{k_e}(m)$ or $E(m)$ for $E(1^{\sigma}, k_e, m)$ and $D_k(c)$ or $D(c)$ for $D(1^{\sigma}, k, c)$ when there is no possibility of any ambiguity. If the scheme is probabilistic, we will also write $E_{k_e}(m)$ or $E(m)$ as well as $E_{k_e}(m, r)$ or $E(m, r)$ for $E(1^{\sigma}, k_e, m, r)$. Furthermore, we will write $A\big(E(m), E(m')\big) = E(m \, o \, m')$ to denote that the algorithm $A$ (either *Add* or *Mult*) is applied on two encryptions of the messages $m, m' \in (\boldsymbol{M}, o)$ and outputs an encryption of $m \, o \, m'$, i.e., it holds that except with negligible probability:

$$D\left(A\left(1^{\sigma}, k_e, E_{k_e}(m), E_{k_e}(m')\right)\right) = m \, o \, m'$$

**Example:** In the following, we give an example of a deterministic multiplicatively homomorphic scheme and an example of a probabilistic, additively homomorphic scheme.

**The RSA Scheme:** The classical RSA scheme (Rivest et al., 1987b) is an example of a deterministic multiplicatively homomorphic cryptosystem on $\boldsymbol{M} = (\mathbb{Z}/N\mathbb{Z}, .)$, where N is the product of two large primes. As ciphertext space, we have $\boldsymbol{C} = (\mathbb{Z}/N\mathbb{Z}, .)$ and as key space we have $\boldsymbol{K} = \big\{(k_e, k_d) = \big((N, e), d\big) \,\big|\, N = pq, ed \equiv 1 \, mod \, \varphi(N)\big\}$. The encryption of a message $m \in \boldsymbol{M}$ is defined as $E_{k_e}(m) = m^e \, mod \, N$ for decryption of a ciphertext $E_{k_e}(m) = c \in \boldsymbol{C}$ we compute $D_{k_e, k_d}(c) = c^d \, mod \, N = m \, mod \, N$. Obviously, the encryption of the product of two messages can be efficiently computed by multiplying the corresponding ciphertexts, i.e.,

$$E_{k_e}(m_1 . m_2) = (m_1 . m_2)^e \, mod \, N = (m_1^e \, mod \, n)(m_2^e \, mod \, N) = E_{k_e}(m_1) . E_{k_e}(m_2)$$

where $m_1, m_2 \in \boldsymbol{M}$. Therefore, the algorithm for *Mult* can be easiliy realized as follows:

$$Mult\left(E_{k_e}(m_1), E_{k_e}(m_2)\right) = E_{k_e}(m_1) . E_{k_e}(m_2)$$

Usually in the RSA scheme as well as in most of the cryptosystems which are based on the difficulty of factoring the security parameter σ is the bit length of $N$. For instance, σ = 1024 is a common security parameter.

**The Goldwasser-Micali Scheme:** The Goldwasser-Micali scheme (Goldwasser & Micali, 1984) is an example of a probabilistic, additively homomorphic cryptosystem on $\boldsymbol{M} = (\mathbb{Z}/2\mathbb{Z}, +)$ with the ciphertext space $\boldsymbol{C} = \boldsymbol{Z} = (\mathbb{Z}/N\mathbb{Z})^*$ where $N = pq$ is the product of two large primes. We have.

$$K = \{(k_e, k_d) = ((N, a), (p, q)) \mid N = pq, \ a \in (\mathbb{Z}/N\mathbb{Z})^* : \left(\frac{a}{p}\right) = \left(\frac{a}{q}\right) = -1\}$$

Since this scheme is probabilistic, the encryption algorithm gets as additional input a random value $r \in \boldsymbol{Z}$. We define $E_{k_e}(m, r) = a^m r^2 \bmod N$ and $D_{(k_e k_d)} = 0$ if $c$ is a square and = 1 otherwise. The following relation therefore holds good:

$$E_{k_e}(m_1, r_1) . E_{k_e}(m_2, r_2) = E_{k_e}(m_1 + m_2, r_1 r_2)$$

The algorithms *Add* can, therefore, be efficiently implemented as follows:

$$Add\big(E_{k_e}(m_1, r_1), E_{k_e}(m_2, r_2), r_3\big) = E_{k_e}(m_1, r_1). E_{k_e}(m_2, r_2). r_3^2 \bmod N = E_{ke}(m_1 + m_2, \ r_1 r_2 r_3)$$

In the above equation, $r_3^2 \bmod N$ is equivalent to $E_{k_e}(0, r_3)$. Also, $m_1, m_2 \in \boldsymbol{M}$ and $r_1, r_2, r_3 \in \boldsymbol{Z}$. Note that this algorithm should be probabilistic, since it obtains a random number $r_3$ as an additional input.

A public-key homomorphic encryption scheme on a (semi-)ring $(\boldsymbol{M}, +, .)$ can be defined in a similar manner. Such schemes consist of two algorithms: *Add* and *Mult* for the homomorphic property instead of one algorithm for *A*, i.e., it is additively and multiplicatively homomorphic at the same time. Such schemes are called *algebraically homomorphic*.

**Definition:** An additively homomorphic encryption scheme on a (semi-)ring $(\boldsymbol{M}, +, .)$ is called *scalar homomorphic* if there exists a probabilistic, expected polynomial time algorithm *Mixed_Mult* that on inputs $1^\sigma, k_e, s \in \boldsymbol{M}$ and an element $c \in \boldsymbol{C}$ outputs an element $c' \in \boldsymbol{C}$ so that for all $m \in \boldsymbol{M}$ it holds that: if $m' = s.m$ and $c = E(1^\sigma, k_e, m)$ then the probability $Prob\big[D\big(Mixed\_Mult(1^\sigma, k_e, s, s)\big) \neq m'\big]$ is negligible.

Thus in a scalar homomorphic scheme, it is possible to compute an encryption $E(1^\sigma, k_e, s.m) = E(1^\sigma, k_e, m')$ of a product of two messages $s, m \in \boldsymbol{M}$ given the public key $k_e$ and an encryption $c = E(1^\sigma, k_e, m)$ of one message $m$ and the other message $s$ as a plaintext. It is clear that any scheme that is algebraically homomorphic is scalar homomorphic as well.

We will denote by $Mixed\_Mult\big(m, E(m')\big) = E(mm')$ if the following equation holds good except possibly with a negligible probability of not holding.

$$D\big(Mixed\_Mult\left(1^\sigma, k_e, m, E_{k_e}(m')\right) = m . m'$$

**Definition:** A *blinding algorithm* is a probabilistic, polynomial-time algorithm which on inputs $1^\sigma, k_e$, and $c \in E_{k_e}(m, r)$ where $r \in \mathbb{Z}$ is randomly chosen outputs another encryption $c' \in E_{k_e}(m, r')$ of $m$ where $r' \in \mathbb{Z}$ is chosen uniformly at random.

For instance, in a probabilistic, homomorphic cryptosystem on $(\boldsymbol{M}, o)$ the blinding algorithm can be realized by applying the algorithm *A* on the ciphertext *c* and an encryption of the identity element in $\boldsymbol{M}$.

Homomorphic Encryption: Theory and Application

If $M$ is isomorphic to $\mathbb{Z}/n\mathbb{Z}$ if M is finite or to $\mathbb{Z}$ otherwise, then the algorithm *Mixed_Mult* can easily be implemented using a double and *Add* algorithm. This is combined with a blinding algorithm is the scheme is probabilistic (Cramer et al., 2000). Hence, every additively homomorphic cryptosystem on $\mathbb{Z}/n\mathbb{Z}$ or $\mathbb{Z}$ is also *scalar homomorphic* and the algorithm *Mixed_Mult* can be efficiently implemented (Sander & Tschudin, 1998).

**Algebraically Homomorphic Cryptosystems:** The existence of an efficient and secure algebraically homomorphic cryptosystem has been a long standing open question. In this Section, we first present some related work considering this problem. Thereafter, we describe the relationship between algebraically homomorphic schemes and homomorphic schemes on special non-abelian groups. More precisely, we prove that a homomorphic encryption scheme on the non-ableain group ($\mathbf{S_7}$, .), the symmetric group on seven elements, allows to construct an algebraically homomorphic encryption scheme on ($\mathbf{F_2}$, $\mathbf{+}$, .). An algebraically homomorphic encryption scheme on ($\mathbf{F_2}$, $\mathbf{+}$, .) can also be obtained from a homomorphic encryption scheme on the special linear group ($SL(\mathbf{3}, \mathbf{2})$, .) over $\mathbf{F_2}$. Furthermore, using coding theory, an algebraically homomorphic encryption on an arbitrary finite ring or field could be obtained given a homomorphic encryption scheme on one of these non-abelian groups. These observations could be a first step to solve the problem whether efficient and secure algebraically homomorphic schemes exist. The research community in cryptography has spent substantial effort on this problem. In 1996, Boneh and Lipton proved that under a reasonable assumption every deterministic, algebraically homomorphic cryptosystem can be broken in sub-exponential time (Boneh & Lipton, 1996). This may be perceived as a negative result concerning the existence of an algebraically homomorphic encryption scheme, although most of the existing cryptosystems, e.g., RSA scheme or the ElGamal scheme can be also broken in sub-exponential time. Furthermore, if we seek for algebraically homomorphic public-key schemes on small fields or rings such as $M = \mathbf{F_2}$, obviously such a scheme has to be probabilistic in order to be secure.

Some researchers also tried to find candidates for algebraically homomorphic schemes. In 1993, Fellows and Koblitz presented an algebraic public-key cryptosystem called Polly Cracker (Fellows & Koblitz, 1993). It is algebraically homomorphic and provably secure. Unfortunately, the scheme has a number of difficulties and is not efficient concerning the ciphertext length. Firstly, Polly Cracker is a polynomial-based system. Therefore, computing an encryption of the product $E(m_1 . m_2)$ of two messages $m_1$ and $m_2$ by multiplying the corresponding ciphertext polynomials $E(m_1)$ and $E(m_2)$, leads to an exponential blowup in the number of monomials. Hence, during repeated computations, there is an exponential blow up in the ciphertext length. Secondly, all existing instantiations of Polly Cracker suffer from further drawbacks (Koblitz, 1998). They are either insecure since they succumb to certain attacks, they are too inefficient to be practical, or they lose the algebraically homomorphic property. Hence, it is far from clear how such kind of schemes could be turned into efficient and secure algebraically homomorphic encryption schemes. A detailed analysis and description of these schemes can be found in (Ly, 2002).

In 2002, J. Domingo-Ferrer developed a probabilistic, algebraically homomorphic secret-key cryptosystem (Domingo-Ferrer, 2002). However, this scheme was not efficient since there was an exponential blowup in the ciphertext length during repeated multiplications that were required to be performed. Moreover, it was also broken by Wagner and Bao (Bao, 2003; Wagner, 2003).

Thus considering homomorphic encryption schemes on groups instead of rings seems more promising to design a possible algebraically homomorphic encryption scheme. It brings us closer to structures that have been successfully used in cryptography. The following theorem shows that indeed the search for algebraically homomorphic schemes can be reduced to the search for homomorphic schemes on special non-abelian groups (Rappe, 2004).

**Theorem I:** The following two statements are equivalent: (1) There exists an algebraically homomorphic encryption scheme on $(\mathbf{F_2}, +, .)$. (2) There exists a homomorphic encryption scheme on the symmetric group $(\mathbf{S_7}, .)$.

**Proof: 1 ➔ 2:** This direction of proof follows immediately and it holds for an arbitrary finite group since operations of finite groups can always be implemented by Boolean circuits. Let $S_7$ be represented as a subset of $\{0, 1\}^l$, where e.g. $l = 21$ can be chosen, and let $C$ be a circuit with addition and multiplication gates that takes as inputs the binary representations of elements $m_1, m_2 \in S_7$ and outputs the binary representations of $m_1 m_2$. If we have an algebraically homomorphic encryption scheme $(K, E, D, Add, Mult)$ on $(\mathbf{F_2}, +, .)$ then we can define a homomorphic encryption scheme $(\widetilde{K}, \widetilde{E}, \widetilde{D}, \widetilde{Mult})$ on $\mathbf{S_7}$ by defining $\widetilde{E}(m) = (E(s_0), \dots . E(s_{l-1}))$ where $(s_0, \dots \dots . s_{l-1})$ denotes the binary representation of $m$. $\widetilde{Mult}$ is constructed by substituting the addition gates in $C$ by $Add$ and the multiplication gates by $Mult$. $\widetilde{K}$ and $\widetilde{D}$ are defined in the obvious way.

**2 ➔ 1:** The proof has two steps. First, we use a construction of Ben-Or and Cleve (Ben-Or & Cleve, 1992) to show that the field $(\mathbf{F_2}, +, .)$ can be encoded in the special linear group $(\mathbf{SL}(3,2), .)$ over $\mathbf{F_2}$. Then, we apply a theorem from projective geometry to show that $(\mathbf{SL}(3,2), .)$ is a subgroup of $\mathbf{S_7}$. This proves the claim.

Homomorphic encryption schemes on groups have been extensively studied. For instance, we have homomorphic schemes on groups $(\mathbb{Z}/M\mathbb{Z}, +)$, for $M$ being a smooth number (Goldwasser & Micali, 1984; Benaloh, 1994; Naccache & Stern, 1998) for $M = p.q$ being an RSA modulus (Paillier, 1999; Galbraith, 2002), and for groups $((\mathbb{Z}/N\mathbb{Z})^*, .)$ where $N$ is an RSA modulus. All known efficient and secure schemes are homomorphic on abelian groups. However, $S_7$ and $SL(3, 2)$ are non-abelian. Sander, Young and Yung (Sander et al., 1999) investigated the possibility of existence of a homomorphic encryption scheme on non-abelain groups. Although non-abelian groups had been used to construct encryption schemes (Ko et al., 2000; Paeng et al., 2001; Wagner & Magyarik, 1985; Grigoriev & Ponomarenko, 2006), the resulting schemes are not homomorphic in the sense that we need for computing efficiently on encrypted data.

Grigoriev and Ponomarenko propose a novel definition of homomorphic cryptosystems on which they base a method to construct homomorphic cryptosystems over arbitrary finite groups including non-abelian groups (Grigoriev & Ponomarenko, 2006). Their construction method is based on the fact that every finite group is an epimorphic image of a free product of finite group cyclic groups. It uses existing homomorphic encryption schemes on finite cyclic groups as building blocks to obtain homomorphic encryption schemes on arbitrary finite groups. Since the ciphertext space obtained from the encryption scheme is a free product of

Homomorphic Encryption: Theory and Application

groups, an exponential blowup of the ciphertext lengths during repeated computations is produced as a result. The reason is that the length of the product of two elements $x$ and $y$ of a free product is, in general, the sum of the length of $x$ and the length of $y$. Hence, the technique proposed by Grigoriev and Ponomarenko suffers from the same drawback as the earlier schemes and does not provide an efficient cryptosystem. We note that using this construction it is possible to construct a homomorphic encryption scheme on the symmetric group $\mathbf{S_7}$ and on the special linear group $SL(3, 2)$. If we combine this with **Theorem 1**, we can construct an algebraically homomorphic cryptosystem on the finite field $(\mathbf{F_2}, \mathbf{+}, .)$. Unfortunately, the exponential blowup owing to the construction method in the homomorphic encryption scheme on $\mathbf{S_7}$ and on $SL(3, 2)$ respectively, would lead to an exponential blowup in $\mathbf{F_2}$ and hence leaves the question open if an efficient algebraically homomorphic cryptosystem on $\mathbf{F_2}$ exists. We will come back to this issue in Section 6, where we discuss *fully homomorphic encryption schemes*.

Grigoriev and Ponomarenko propose another method to encrypt arbitrary finite groups homomorphically (Grigoriev & Ponomarenko, 2004). This method is based on the difficulty of the membership problem for groups of integer matrices, while in (Grigoriev & Ponomarenko, 2006) it is based on the difficulty of factoring. However, as before, this scheme is not efficient. Moreover, in (Grigoriev & Ponomarenko, 2004), an algebraically homomorphic cryptosystem over finite commutative rings is proposed. However, owing to its immense size, it is infeasible to implement in real-world applications.

## 4. Some Classical Homomorphic Encryption Systems

In this Section, we describe some classical homomorphic encryption systems which have created substantial interest among the researchers in the domain of cryptography. We start with the first probabilistic systems proposed by Goldwasser and Micali in 1982 (Goldwasser & Micali, 1982; Goldwasser & Micali, 1984) and then discuss the famous Paillier's encryption scheme (Paillier, 1999) and its improvements. Paillier's scheme and its variants are well-known for their efficiency and the high level of security that they provide for homomorphic encryption. We do not discuss their mathematical considerations in detail, but summarize their important parameters and properties.

**Goldwasser-Micali scheme:** This scheme (Goldwasser & Micali, 1982; Goldwasser & Micali, 1984) is historically very important since many of subsequent proposals on homomorphic encryption were largely motivated by its approach. Like in RSA, in this scheme, we use computations modulo $n = p.q$, a product of two large primes. The encryption process is simple which uses a product and a square, whereas decryption is heavier and involves exponentiation. The complexity of the decryption process is: $O(k.l(p)^2)$, where $l(p)$ denotes the number of bits in $p$. Unfortunately, this scheme has a limitation since its input consist of a single bit. First, this implies that encrypting $k$ bits leads to a cost of $O(k.l(p)^2)$. This is not very efficient even if it may be considered as practical. The second concern is related to the issue of *expansion* – a single bit of plaintext is encrypted in *an integer modulo n*, that is, $l(n)$ bits. This leads to a huge blow up of ciphertext causing a serious problem with this scheme.

Goldwasser-Micali (GM) scheme can be viewed from another perspective. When looked from this angle, the basic principle of this scheme is to partition a well-chosen subset of integers modulo $n$ into two secret parts: $M_0$ and $M_1$. The encryption process selects a random element $M_b$ to encrypt plaintext $b$, and the decryption process lets the user know in which part the randomly selected element lies. The essence of the scheme lies in the mechanism to determine the subset, and to partition it into $M_0$ and $M_1$. The scheme uses group theory to achieve this goal. The subset is the group $G$ of invertible integers modulo $n$ with a Jacobi symbol with respect to $n$, equal to 1. The partition is generated by another group $H \subset G$, consisting of the elements that are invertible modulo $n$ with a Jacobi symbol, with respect to a fixed factor $n$, equal to 1. With these settings of parameters, it is possible to split $G$ into two parts – $H$ and $G \backslash H$. The generalization schemes of GM deal with these two groups. These schemes attempt to find two groups $G$ and $H$ such that $G$ can be split into more than $k = 2$ parts.

**Benaloh's scheme:** Benaloh (Benaloh, 1988) is a generalization of GM scheme that enables one to manage inputs of $l(k)$ bits, $k$ being a prime satisfying some specified constraints. Encryption is similar as in GM scheme (encrypting a message $m \in \{0, \dots, k-1\}$ is tantamount to picking an integer $r \in Z_n^*$ and computing $c = g^m r^k \bmod n$). However, the decryption phase is more complex. If the input and output sizes are $l(k)$ and $l(n)$ bits respectively, the expansion is equal to $l(n)/l(k)$. The value of expansion obtained in this approach is less than that achieved in GM. This makes the scheme more attractive. Moreover, the encryption is not too expensive as well. The overhead in the decryption process is estimated to be $O(\sqrt{k}.l(k))$ for pre-computation which remains constant for each dynamic decryption step. This implies that the value of $k$ has to be taken very small, which in turn limits the gain obtained on the value of expansion.

**Naccache-Stern scheme:** This scheme (Naccache & Stern, 1998) is an improvement of Benaloh's scheme. Using a value of the parameter $k$ that is greater than that used in the Benaloh's scheme, it achieves a smaller expansion and thereby attains a superior efficiency. The encryption step is precisely the same as in Benaloh's scheme. However, decryption is different. The value of expansion is same as that in Benaloh's scheme, i.e., $l(n)/l(k)$. However, the cost of decryption is less and is given by: $O(l(n)^5 \log(l(n)))$. The authors claim that it is possible to choose the values of the parameters in the system in such a way that the achieved value of expansion is 4 (Naccache & Stern, 1998).

**Okamoto-Uchiyama scheme:** To improve the performance of the earlier schemes on homomorphic encryption, Okamoto and Uchiyama changed the base group $G$ (Okamoto & Uchiyama, 1998). By taking $n = p^2 q$, $p$ and $q$ being two large prime numbers as usual, and the group $G = Z_{p^2}^*$, the authors achieve $k = p$. The value of the expansion obtained in the scheme is 3. One of the biggest advantages of this scheme is that its security is equivalent to the factorization of $n$. However, a chosen-ciphertext attack has been proposed on this scheme that can break the factorization problem. This has somewhat limited is applicability. However, this scheme was used to design the EPOC systems (Okamoto et al., 2000) which is accepted in the *IEEE standard specifications for public-key cryptography* (IEEE P1363).

**Paillier scheme:** One of the most well-known homomorphic encryption schemes is due to Paillier (Paillier, 1999). It is an improvement over the earlier schemes in the sense that it is able to decrease the value of expansion from 3 to 2. The scheme uses $n = p.q$ with $\gcd(n, \phi(n)) = 1$. As usual $p$ and $q$ are two large primes. However, it considered the group $G = Z_{n^2}^*$ and a proper choice of $H$ led to $k = l(n)$. While the cost of encryption is not

Homomorphic Encryption: Theory and Application

too high, decryption needs one exponentiation modulo $n^2$ to the power $\lambda(n)$, and a multiplication modulo $n$. This makes decryption a bit heavyweight process. The author has shown how to manage decryption efficiently using the famous *Chinese Remainder Theorem*. With smaller expansion and lower cost compared with the other schemes, this scheme found great acceptance. In 2002, Cramer and Shoup proposed a general approach to achieve higher security against *adaptive chosen-ciphertext attacks* for certain cryptosystems with some particular algebraic properties (Cramer & Shoup, 2002). They applied their propositions on Paillier's original scheme and designed a stronger variant of homomorphic encryption. Bresson et al. proposed a slightly different version of a homomorphic encryption scheme that is more accurate for some applications (Bresson et al., 2003).

**Damgard-Jurik scheme:** Damgard and Jurik propose a generalization of Paillier's scheme to groups of the form $\mathbf{Z}^*_{n^{s+1}}$ for $s > 0$ (Damgard & Jurik, 2001). In this scheme, choice of larger values of $s$ will achieve lower values of expansion. This scheme can be used in a number of applications. For example, we can mention the adaptation of the size of the plaintext, the use of threshold cryptography, electronic voting, and so on. To encrypt a message, $m \in \mathbf{Z}^*_{n^s}$, one picks at random $r \in \mathbf{Z}^*_n$ and computes $g^m r^{n^s} \in \mathbf{Z}_{n^{s+1}}$. The authors show that if one can break the scheme for a given value $s = \sigma$, then one can break it for $s = \sigma - 1$. They also show that the semantic security of this scheme is equivalent to that of Paillier's scheme. The value of expansion can be computed using: $1 + 1/s$. It is clear that expansion can attain a value close to 1 if $s$ is sufficiently large. The ratio of the cost for encryption in this scheme over Paillier's scheme can be estimated to be: $\frac{s(s+1)(s+2)}{6}$. The same ratio for the decryption process will have value equal to: $\frac{(s+1)(s+2)}{6}$. Even if this scheme has a lower value of expansion as compared to Paillier's scheme, it is computationally more intensive. Moreover, if we want to encrypt or decrypt $k$ blocks of $l(n)$ bits, running Paillier's scheme $k$ times is less expensive than running Damgard-Jurik's scheme.

**Galbraith scheme:** This is an adaptation of the existing homomorphic encryption schemes in the context of elliptic curves (Galbraith, 2002). Its expansion is equal to 3. For $s = 1$, the ratio of the encryption cost for this scheme over that of Paillier's scheme can be estimated to be about 7, while the same ratio for the cost of decryption cost is about 14 for the same value of $s$. However, the most important advantage of this scheme is that the cost of encryption and decryption can be decreased using larger values of $s$. In addition, the security of the scheme increases with the increase in the value of $s$ as it is the case in Damgard-Jurik's scheme.

**Castagnos scheme:** Castagnos explored the possibility of improving the performance of homomorphic encryption schemes using quadratic fields quotations (Castagnos, 2006; Castagnos, 2007). This scheme achieves an expansion value of 3 and the ratio of encryption/decryption cost with $s = 1$ over Paillier's scheme can be estimated to be about 2.

## 5. Applications and Properties of Homomorphic Encryption Schemes

An inherent drawback of homomorphic cryptosystems is that attacks on these systems might possibly exploit their additional structural information. For instance, using plain RSA (Rivest et al., 1978b) for signing, the multiplication of two signatures yields a valid signature of the product of the two corresponding messages. Although there are many ways to avoid such attacks, for instance, by application of hash functions, the use of redundancy or probabilistic schemes, this potential weakness leads us to the question why homomorphic schemes should be used instead of conventional cryptosystems under certain situations. The main reason for the interest in homomorphic cryptosystems is its wide application scope. There are theoretical as well as practical applications in different areas of cryptography. In the following, we list some of the main applications and properties of homomorphic schemes and summarize the idea behind them.

### 5.1 Some Applications of Homomorphic Encryption Schemes

**Protection of mobile agents:** One of the most interesting applications of homomorphic encryption is its use in protection of mobile agents. As we have seen in Section 3, a homomorphic encryption scheme on a special non-abelian group would lead to an algebraically homomorphic cryptosystem on the finite filed $F_2$. Since all conventional computer architectures are based on binary strings and only require multiplication and addition, such homomorphic cryptosystems would offer the possibility to encrypt a whole program so that it is still executable. Hence, it could be used to protect mobile agents against malicious hosts by encrypting them (Sander & Tschudin, 1998a). The protection of mobile agents by homomorphic encryption can be used in two ways: (i) *computing with encrypted functions* and (ii) *computing with encrypted data*. Computation with encrypted functions is a special case of protection of mobile agents. In such scenarios, a secret function is publicly evaluated in such a way that the function remains secret. Using homomorphic cryptosystems the encrypted function can be evaluated which guarantees its privacy. Homomorphic schemes also work on encrypted data to compute publicly while maintaining the privacy of the secret data. This can be done encrypting the data in advance and then exploiting the homomorphic property to compute with encrypted data.

**Multiparty computation:** In multiparty computation schemes, several parties are interested in computing a common, public function on their inputs while keeping their individual inputs private. This problem belongs to the area of *computing with encrypted data*. Usually in multiparty computation protocols, we have a set of $n \geq 2$ players whereas in computing with encrypted data scenarios $n = 2$. Furthermore, in multi-party computation protocols, the function that should be computed is publicly known, whereas in the area of computing with encrypted data it is a private input of one party.

**Secret sharing scheme:** In secret sharing schemes, parties share a secret so that no individual party can reconstruct the secret form the information available to it. However, if some parties cooperate with each other, they may be able to reconstruct the secret. In this scenario, the homomorphic property implies that the composition of the shares of the secret is equivalent to the shares of the composition of the secrets.

**Threshold schemes:** Both secret sharing schemes and the multiparty computation schemes are examples of threshold schemes. Threshold schemes can be implemented using homomorphic encryption techniques.

**Zero-knowledge proofs:** This is a fundamental primitive of cryptographic protocols and serves as an example of a theoretical application of homomorphic cryptosystems. Zero-knowledge proofs are used to prove knowledge of some private information. For instance,

Homomorphic Encryption: Theory and Application

consider the case where a user has to prove his identity to a host by logging in with her account and private password. Obviously, in such a protocol the user wants her private information (i.e., her password) to stay private and not to be leaked during the protocol operation. Zero-knowledge proofs guarantee that the protocol communicates exactly the knowledge that was intended, and no (zero) extra knowledge. Examples of zero-knowledge proofs using homomorphic property can be found in (Cramer & Damgard, 1998).

**Election schemes:** In election schemes, the homomorphic property provides a tool to obtain the tally given the encrypted votes without decrypting the individual votes.

**Watermarking and fingerprinting schemes:** Digital watermarking and fingerprinting schemes embed additional information into digital data. The homomorphic property is used to add a mark to previously encrypted data. In general, watermarks are used to identify the owner/seller of digital goods to ensure the copyright. In fingerprinting schemes, the person who buys the data should be identifiable by the merchant to ensure that data is not illegally redistributed. Further properties of such schemes can be found in (Pfitzmann & Waidner, 1997; Adelsbach et al. 2002).

**Oblivious transfer:** It is an interesting cryptographic primitive. Usually in a two-party 1-out-of-2 oblivious transfer protocol, the first party sends a bit to the second party in such as way that the second party receives it with probability ½, without the first party knowing whether or not the second party received the bit. An example of such a protocol that uses the homomorphic property can be found in (Lipmaa, 2003).

**Commitment schemes:** Commitment schemes are some fundamental cryptographic primitives. In a commitment scheme, a player makes a commitment. She is able to choose a value from some set and commit to her choice such that she can no longer change her mind. She does not have to reveal her choice although she may do so at some point later. Some commitment schemes can be efficiently implemented using homomorphic property.

**Lottery protocols:** Usually in a cryptographic lottery, a number pointing to the winning ticket has to be jointly and randomly chosen by all participants. Using a homomorphic encryption scheme this can be realized as follows: Each player chooses a random number which she encrypts. Then using the homomorphic property the encryption of the sum of the random values can be efficiently computed. The combination of this and a *threshold decryption scheme* leads to the desired functionality. More details about homomorphic properties of lottery schemes can be found in (Fouque et al., 2000).

**Mix-nets:** Mix-nets are protocols that provide anonymity for senders by collecting encrypted messages from several users. For instance, one can consider mix-nets that collect ciphertexts and output the corresponding plaintexts in a randomly permuted order. In such a scenario, privacy is achieved by requiring that the permutation that matches inputs to outputs is kept secret to anyone except the mix-net. In particular, determining a correct input/output pair, i.e., a ciphertext with corresponding plaintext, should not be more effective then guessing one at random. A desirable property to build such mix-nets is re-encryption which is achieved by using homomorphic encryption. More information about applications of homomorphic encryption in mix-nets can be found in (Golle et al., 2004; Damgard & Jurik, 2003).

**Data aggregation in wireless sensor networks:** In-network data aggregation in WSNs is a technique that combines partial results at the intermediate nodes en route to the base station (i.e., the node issuing the query), thereby reducing the communication overhead and optimizing the bandwidth utilization in the wireless links (Sen 2012a; Sen & Maitra, 2011;). However, this technique raises privacy and security issues if the sensor nodes which need to share their data with the aggregator node. In applications such as healthcare and military surveillance where the sensitivity of private data of the sensor is very high, the aggregation has to be carried out in a privacy-preserving way, so that the sensitive data are not revealed to the aggregator. Secure and efficient schemes for additive data aggregation functions have been proposed in (Sen 2012a; Sen & Maitra 2011; He et al., 2007). However, these schemes are not generic and cannot be applied for multiplicative aggregation function. Homomorphic encryption schemes can be applied to protect privacy of input data while computing an arbitrary aggregation function in a wireless sensor network. Security in sensor networks is a well researched problem (Sen 2012b; Sen & Ukil, 2010; Sen, 2010a; Sen 2010b; Sen, 2009). Security in data aggregation has also been studied extensively by researchers. Interested readers may refer to (Sen, 2012a; Sen 2011a) for further details.

### 5.2 Some Properties of Homomorphic Encryption Schemes

Homomorphic encryption schemes have some interesting mathematical properties. In the following, we mention some of these properties.

**Re-randomizable encryption/re-encryption:** Re-randomizable cryptosystems (Groth, 2004) are probabilistic cryptosystems with the additional property that given the public key $k_e$ and an encryption $E_{k_e}(m, r)$ of a message $m \in M$ under the public key $k_e$ and a random number $r \in Z$ it is possible to efficiently convert $E_{k_e}(m, r)$ into another encryption $E_{k_e}(m, r')$ that is perfectly indistinguishable from a *fresh* encryption of $m$ under the public key $k_e$. This property is also called *re-encryption*.

It obvious that every probabilistic homomorphic cryptosystem is re-randomizable. Without loss of generality, we assume that the cryptosystem is additively homomorphic. Given $E_{k_e}(m, r)$ and the public key $k_e$, we can compute $E_{k_e}(0, r'')$ for a random number $r''$ and hence compute the following:

$$Add\left(E_{k_e}(m, r), E_{k_e}(0, r'')\right) = E_{k_e}(m + 0, r') = E_{k_e}(m, r')$$

where $r'$ is an appropriate random number. We note that this is exactly what a *blinding algorithm* does.

**Random self-reducibility:** Along with the possibility of re-encryption comes the property of random self-reducibility concerning the problem of computing the plaintext from the ciphertext. A cryptosystem is called *random self-reducible* if any algorithm that can break a non-trivial fraction of ciphertexts can also break a random instance with significant probability. This property is discussed in detail in (Damgard et al., 2010; Sander et al., 1999).

**Verifiable encryptions / fair encryptions:** If an encryption is verifiable, it provides a mechanism to check the correctness of encrypted data without compromising on the secrecy of the data. For instance, this is useful in voting schemes to convince any observer that the

Homomorphic Encryption: Theory and Application

encrypted name of a candidate, i.e., the encrypted vote is indeed in the list of candidates. A cryptosystem with this property that is based on homomorphic encryption can be found in (Poupard & Stern, 2000). Verifiable encryptions are also called *fair encryptions*.

## 6. Fully Homomorphic Encryption Schemes

In 2009, Gentry described the first plausible construction of a fully homomorphic cryptosystem that supports both addition and multiplication (Gentry, 2009). Gentry's proposed fully homomorphic encryption consists of several steps: First, it constructs a *somewhat homomorphic* scheme that supports evaluating low-degree polynomials on the encrypted data. Next, it *squashes* the decryption procedure so that it can be expressed as a low-degree polynomial which is supported by the scheme, and finally, it applies a *bootstrapping transformation* to obtain a fully homomorphic scheme. The essential approach of this scheme is to derive and establish a process that can evaluate polynomials of high-enough degree using a decryption procedure that can be expressed as a polynomial of low-enough degree. Once the degree of polynomials that can be evaluated by the scheme exceeds the degree of the decryption polynomial by a factor of two, the scheme is called *bootstrappable* and it can then be converted into a fully homomorphic scheme.

For designing a bootstrappable scheme, Gentry presented a somewhat homomorphic scheme (Gentry, 2009) which is roughly a GGH (Goldreich, Goldwasser, Halevi)-type scheme (Goldreich et al., 1997; Micciancio, 2001) over ideal lattices. Gentry later proved that with an appropriate key-generation procedure, the security of that scheme can be reduced to the worst-case hardness of some lattice problems in ideal lattice constructions (Gentry, 2010). Since this somewhat homomorphic scheme is not bootstrappable, Gentry described a transformation to squash the decryption procedure, reducing the degree of the decryption polynomial (Gentry, 2009). This is done by adding to the public key, an additional hint about the secret key in the form of a *sparse subset-sum problem* (SSSP). The public key is augmented with a big set of vectors in such a way that there exists a very sparse subset of them that adds up to the secret key. A ciphertext of the underlying scheme can be *post-processed* using this additional hint and the post-processed ciphertext can be decrypted with a low-degree polynomial, thereby achieving a bootstrappable scheme.

Gentry's construction is quite involved – the secret key, even in the private key version of his scheme is a short basis of a *random ideal lattice*. Generating pairs of public and secret bases with the right distributions appropriate for the worst-case to average-case reduction is technically quite complicated. A significant research effort has been devoted to increase the efficiency of its implementation (Gentry & Halevi, 2011; Smart & Vercauteren, 2010).

A parallel line of work that utilizes ideal lattices in cryptography dates back to the NTRU cryptosystem (Hoffstein et al., 1998). This approach uses ideal lattices for efficient cryptographic constructions. The additional structure of ideal lattices, compared to ordinary lattices, makes their representation more powerful and enables faster computation. Motivated by the work of Micciancio (Micciancio, 2007), a significant number of work (Peikert & Rosen, 2006; Lyubashevsky & Micciancio, 2006; Peikert & Rosen, 2007; Lyubashevsky et al., 2008; Lybashevsky & Micciancio, 2008) has produced efficient constructions of various cryptographic primitives whose security can formally be reduced to the hardness of short-vector problems in ideal lattices (Brakerski & Vaikuntanathan, 2011).

Lyubashevsky et al. (Lyubashevsky et al., 2010) present the *ring learning with errors* (RLWE) assumption which is the *ring counterpart* of Regev's learning with errors assumption (Regev, 2005). In a nutshell, the assumption is that given polynomially many samples over a certain ring of the form $(a_i, a_i s + e_i)$, where $s$ is a random *secret ring element*, $a_i$'s are distributed uniformly randomly in the ring, and $e_i$ are *small* ring elements, it will be impossible for an adversary to distinguish this sequence of samples from random pairs of ring elements. The authors have shown that this simple assumption can be very efficiently reduced to the worst case hardness of short-vector problems on ideal lattices. They have also shown how to construct a very efficient ring counterpart to Regev's public-key encryption scheme (Regev, 2005), as well as a counterpart to the identity-based encryption scheme presented in (Gentry et al., 2008) by using the basis sampling techniques in (Regev, 2005). The scheme presented in (Lyubashevsky et al., 2010) is very elegant and efficient since it is not dependent on any complex computations over ideal lattices.

Brakerski and Vaikuntanathan raised a natural question that whether the above approaches (i.e., ideal lattices and RLWE) can be effectively exploited so that benefits of both these approaches can be achieved at the same time – namely the functional powerfulness on the one hand (i.e., the ideal lattice approach) and the simplicity and efficiency of the other (i.e., RLWE). They have shown that indeed this can be done (Brakerski & Vaikuntanathan, 2011). They have constructed a somewhat homomorphic encryption scheme based on RLWE. The scheme inherits the simplicity and efficiency, as well as the worst case relation to ideal lattices. Moreover, the scheme enjoys *key dependent message security* (KDM security, also known as *circular security*), since it can securely encrypt polynomial functions (over an appropriately defined ring) of its own secret key. The significance of this feature of the scheme in context of homomorphic encryption has been clearly explained by the authors. The authors argue that all known constructions of fully homomorphic encryption employ a bootstrapping technique that enforces the public key of the scheme to grow linearly with the maximal depth of evaluated circuits. This is a major drawback with regard to the usability and the efficiency of the schemes. However, the size of the public key can be made independent of the circuit depth if the somewhat homomorphic scheme can securely encrypt its own secret key. With the design of this scheme, the authors have solved an open problem - achieving *circular secure somewhat homomorphic encryption.* They have also computed the circular security of their scheme with respect to the representation of the secret key as a ring element, where bootstrapping requires circular security with respect to the bitwise representation of the secret key (actually, the bitwise representation of the *squashed* secret key). Since there is no prior work that studies a possible co-existence between somewhat homomorphism with any form of circular security, the work is a significant first step towards removing the assumption (Brakerski & Vaikuntanathan, 2011). The authors have also shown how to transform the proposed scheme into a fully homomorphic encryption scheme following Gentry's blueprint of *squashing* and *bootstrapping.* Applying the techniques presented in (Brakerski & Vaikuntanathan, 2011a), the authors argue that *squashing* can even be avoided at the cost of relying of a *sparse* version of RLWE that is not known to reduce to worst case scenarios. This greatly enhances the efficiency of the proposed scheme in practical applications. The proposed scheme is also *additively key-*

Homomorphic Encryption: Theory and Application

*homomorphic*– a property that has found applications in achieving security against *key-related attacks* (Applebaum et al., 2011).

Smart and Vercauteren (Smart & Vercauteren, 2010) present a fully homomorphic encryption scheme has smaller key and ciphertext sizes. The construction proposed by the authors follows the fully homomorphic construction based on ideal lattices proposed by Gentry (Gentry, 2009). It produces a fully homomorphic scheme form a *somewhat homomorphic scheme*. For somewhat homomorphic scheme, the public and the private keys consist of two large integers (one of which shared by both the public and the private key), and the ciphertext consists of one large integer. The scheme (Smart & Vercauteren, 2010) has smaller *ciphertext blow up* and reduced key size than in Gentry's scheme based on ideal lattices. Moreover, the scheme also allows efficient fully homomorphic encryption over any field of characteristics two. More specifically, it uses arithmetic of *cyclotomic number fields*. In particular, the authors have focused on the field generated by the polynomial: $F(X) = X^{2^n} + 1$. However, they also noted that the scheme could be applied with arbitrary (even non-cyclotomic) number fields as well. In spite of having a large number of advantages, the major problem with this scheme is that the key generation method is very slow.

Gentry and Halevi presented a novel implementation approach for the variant of Smart and Vercauteren proposition (Smart & Vercauteren, 2010), which had a greatly improved key generation phase (Gentry & Halevi, 2011). In particular, the authors have noted that key generation (for cyclotomic fields) is essentially an application of a *Discrete Fourier Transform* (DFT), followed by a small quantum of computation, and then application of the *inverse* transform. The authors then further demonstrate that it is not even required to perform the DFTs if one selects the cyclotomic field to be of the form: $X^{2^n} + 1$. The authors illustrate this by using a recursive approach to deduce two constants from the secret key which subsequently facilitates the key generation algorithm to construct a valid associated public key. The key generation method of Gentry and Halevi (Gentry & Halevi, 2011) is fast. However, the scheme appears particularly tailored to work with two-power roots of unity.

Researchers have also examined ways of improving key generation in fully homomorphic encryption schemes. For example, in (Ogura et al., 2010) a method is proposed for construction of keys for essentially random number fields by pulling random elements and analyzing *eigenvalues* of the corresponding matrices. However, this method is unable to achieve the improvement in efficiency in terms of *reduced ciphertext blow up* as done in (Smart & Vercauteren, 2010) and (Gentry & Halevi, 2011).

Stehle and Steinfield improved Gentry's fully homomorphic scheme and obtained a faster fully homomorphic scheme with $O(n^{3.5})$ bits complexity per elementary binary addition/multiplication gate (Stehle & Steinfield, 2010). However, the hardness assumption of the security of the scheme is stronger than that of Gentry's scheme (Gentry, 2009). The improved complexity of the proposed scheme stems from two sources. First, the authors have given a more aggressive security analysis of the *sparse subset sum problem* (SSSP) against lattice attacks as compared to the analysis presented in (Gentry, 2009). The SSSP along with the ideal lattice *bounded distance decoding* (BDD) problem are the two problems underlying the security of Gentry's fully homomorphic scheme. In his security analysis of BDD, Gentry has used the best known complexity bound for the approximate *shortest vector problem* (SVP) in lattices. However, in analyzing SSSP, Gentry has assumed the availability of an exact SVP oracle. On the contrary, the finer analysis of Stehle and Steinfield for SSSP takes into account the complexity of approximate SVP, thereby making it more consistent with the assumption underlying the analysis of the BDD problem. This leads to choices of smaller parameter in the scheme. Moreover, Stehle and Steinfeld have relaxed the definition of fully homomorphic encryption to allow for a negligible but non-zero probability of decryption error. They have shown that the randomness in the *SplitKey* key generation for the *squashed decryption algorithm* (i.e., the decryption algorithms of the bootstrappable scheme) in the Gentry's scheme can be gainfully exploited to allow a negligible decryption error probability. This decryption error, although negligible in value, can lead to rounding precision used in representing the ciphertext components that is almost half the value of the precision as achieved in Gentry's scheme (Gentry, 2009), which involves zero error probability.

In (Chunsheng, 2012), Chunsheng proposed a modification of the fully homomorphic encryption scheme of Smart and Vercauteren (Smart & Vercauteren, 2010). The author has applied a *self-loop bootstrappable technique* so that the security of the modified scheme only depends on the hardness of the *polynomial coset problem* and does not require any assumption of the *sparse subset problem* as required in the original work of Smart and Vercauteren (Smart & Vercauteren, 2010). In addition, the author have constructed a *non-self-loop fully homomorphic encryption scheme* that uses *cycle keys*. In a nutshell, the security of the improved fully homomorphic encryption scheme in this work is based on use of three mathematical approaches: (i) hardness of factoring integer problem, (ii) solving Diophantine equation problem, and (iii) finding approximate greatest common divisor problem.

Boneh and Freeman propose a linearly homomorphic signature scheme that authenticates vector subspaces of a given ambient space (Boneh & Freeman, 2011). The scheme has several novel features that were not present in any of the existing similar schemes. First, the scheme is the first of its kind that enables *authentication of vectors over binary fields*; previous schemes could not authenticate vectors with large or growing coefficients. Second, the scheme is the only scheme that is based on the *problem of finding short vectors in integer lattices*, and therefore, it enjoys the worst-case security guarantee that is common to *lattice-based cryptosystems*. The scheme can be used to authenticate linear transformations of signed data, such as those arising when computing mean and Fourier transform or in networks that use *network coding* (Boneh & Freeman, 2011). The work has three major contributions in the state of the art as identified by the authors: (i) *Homomorphic signatures over $F_2$:* the authors have constructed the first *unforgeable linearly homomorphic signature scheme* that authenticates vectors with coordinates in $F_2$. It is an example of a cryptographic primitive that can be built using lattice models, but cannot be built using bilinear maps or other traditional algebraic methods based on factoring or discrete log type problems. The scheme can be modified to authenticate vectors with coefficients in other small fields, including prime fields and extension fields such as $F_{2^d}$. Moreover, the scheme is private, in the sense that a derived signature on a vector **v** leaks no information about the original signed vectors beyond what is revealed by **v**. (ii) *A simple k-time signature without random oracles:* the authors have presented a stateless signature scheme and have proved that it is secure in the standard model when used to sign at most $k$ messages, for small values of $k$. The public key of the scheme is significantly smaller than that of any other stateless lattice-based signature scheme that can sign multiple large messages and is secure in the standard model. The construction proposed by the authors can be viewed as *removing the random oracle* from the signature scheme of Gentry, Peikert, and Vaikuntanathan (Gentry et al., 2008), but only for signing $k$ messages (Boneh & Freeman, 2011). (iii) *New tools for lattice-based signatures:* the scheme is unforgeable based on a new hard problem on lattices, which the authors have called the *k-small integer solutions* (*k*-SIS) problem. The authors have shown that *k*-SIS reduces to the *small integer solution* (SIS) problem, which is known to be as hard as standard worst-case lattice problems (Micciancio & Regev, 2007).

## 7. Conclusion and Future Trends

The study of fully homomorphic encryption has led to a number of new and exciting concepts and questions, as well as a powerful tool-kit to address them. We conclude the chapter by discussing a number of research directions related to the domain of fully homomorphic encryption and more generally, on the problem of computing on encrypted data.

**Applications of fully homomorphic encryption:** While Gentry's original construction was considered as being infeasible for practical deployments, recent constructions and implementation efforts have drastically improved the efficiency of fully homomorphic encryption (Vaikuntanathan, 2011). The initial implementation efforts focused on Gentry's original scheme and its variants (Smart & Vercauteren, 2010; Smart & Vercauteren, 2012; Coron et al., 2011; Gentry & Halevi, 2011), which seemed to pose rather inherent efficiency bottlenecks. Later implementations leverage the recent algorithmic advances (Brakerski & Vaikuntanathan, 2011; Brakerski et al., 2011; Brakerski & Vaikuntanathan, 2011a) that result in asymptotically better fully homomorphic encryption systems, as well as new algebraic mechanisms to improve the overall efficiency of these schemes ( Naehrig et al., 2011; Gentry et al., 2012; Smart & Vercauteren, 2012).

**Non-malleability and homomorphic encryption:** Homomorphism and *non-malleability* are two orthogonal properties of an encryption scheme. Homomorphic encryption schemes permit anyone to transform an encryption of a message $m$ into an encryption of $f(m)$ for non-trivial functions $f$. Non-malleable encryption, on the other hand, prevents precisely this sort of thing- it requires that no adversary be able to transform an encryption of $m$ into an encryption of any *related* message. Essentially, what we need is a combination of both the properties that *selectively permit homomorphic computations* (Vaikuntanathan, 2011). This implies that the evaluator should be able to homomorphically compute any function from some pre-specified class $F_{hom}$; however, she should not be able to transform an encryption of $m$ into an encryption of $f(m)$ for which $f \in F_{hom}$ does not hold good (i.e., $f$ does not belong to $F_{hom}$). The natural question that arises is: *whether we can control what is being (homomorphically) computed*?

Answering this question turns out to be tricky. Boneh, Segev and Waters (Boneh et al., 2011) propose the notion of *targeted malleability* – a possible formalization of such a requirement as well as formal constructions of such encryption schemes. Their encryption scheme is based on a strong *knowledge of exponent-type* assumption that allows iterative evaluation of at most $t$ functions, where $t$ is a suitably determined and pre-specified constant. Improving their construction as well as the underlying complexity assumptions is an important open problem (Vaikuntanathan, 2011).

It is also interesting to extend the definition of non-malleability to allow for *chosen cipher-text attacks*. As an example, we consider the problem that involves *implementing an encrypted targeted advertisement system that generates advertisements depending on the contents of a user's e-mail*. Since the e-mail is stored in an encrypted form with the user's public key, the e-mail server performs a homomorphic evaluation and computes an encrypted advertisement to be sent back to the user. The user decrypts it, performs an action depending on what she sees. If the advertisement is relevant, she might choose to click on it; otherwise, she simply discards it. However, if the e-mail server is aware to this information, namely whether the user clicked on the advertisement or not, it can use this as a restricted *decryption oracle* to break the security of the user's encryption scheme and possibly even recover her secret key. Such attacks are ubiquitous whenever we compute on encrypted data, almost to the point that CCA security seems inevitable. Yet, it is easy to see that chosen ciphertext (CCA2-secure) homomorphic encryption schemes cannot exist. Therefore, an appropriate security definition and constructions that achieve the definition is in demand.

**Fully homomorphic encryption and functional decryption:** Homomorphic encryption schemes permit anyone to evaluate functions on encrypted data, but the evaluators never see any information about the result. It is possible to construct an encryption scheme where a user can compute $f(m)$ from an encryption of a message $m$, but she should not be able to learn any other information about $m$ (including the intermediate results in the computation of $f$)? Essentially, the issue boils down to the following question: *can we control the information that the evaluator can see?* Such an encryption scheme is called a *functional encryption scheme.* The concept of functional encryption scheme was first introduced by Sahai and Waters (Sahai & Waters, 2005) and subsequently investigated in a number of intriguing works (Katz et al., 2013; Lewko et al., 2010; Boneh et al., 2011; Agrawal et al., 2011). Although the constructions in these propositions work for several interesting families of functions (such as monotone formulas and inner products), construction of a fully functional encryption scheme is still not achieved and remains as an open problem. What we need is a novel and generic encryption system that provides us with fine-grained control over what one can see and access and what one can compute on data to get a desired output.

**Other problems and applications:** Another important open question relates to the assumptions underlying the current fully homomorphic encryption systems. All known fully homomorphic encryption schemes are based on *hardness of lattice problems*. The natural question that arises - can we construct fully homomorphic from other approaches – say, for example, from number-theoretic assumptions? Can we bring in the issue of the hardness of factoring or discrete logarithms in this problem?

In addition to the scenarios where it is beneficial to keep all data encrypted and to perform computations on encrypted data, fully homomorphic encryption can be gainfully exploited to solve a number of practical problems in cryptography. Two such examples are the problems of *verifiably outsourcing computation* (Goldwasser et al., 2008; Gennaro et al., 2010; Chung et al., 2010; Applebaum et al., 2010) and *constructing short non-interactive zero-knowledge proofs* (Gentry, 2009). Some of the applications of fully homomorphic encryption do not require its full power. For example, in *private information retrieval* (PIR), it is sufficient

Homomorphic Encryption: Theory and Application

to have a somewhat homomorphic encryption scheme that is capable of evaluating simple database indexing functions. For this applications, what is needed is an optimized and less functional encryption scheme that is more efficient than a fully homomorphic encryption function. Design of such functions for different application scenarios is also a current hot topic of research.

Homomorphic Encryption: Theory and Application

Homomorphic Encryption: Theory and Application

Homomorphic Encryption: Theory and Application